\documentclass[letterpaper,10pt]{article} 
\usepackage{osameet3} %% use version 3 for proper copyright statement

\usepackage{amsmath,amssymb}
\usepackage[colorlinks=true,bookmarks=false,citecolor=blue,urlcolor=blue]{hyperref} %pdflatex
\usepackage{subfig}

\begin{document}

\title{Towards Polarization-Insensitive Coherent Coded Phase OTDR}

\author{Sterenn Guerrier$^{1,2}$, Christian Dorize$^1$, Elie Awwad$^{2}$, Jérémie Renaudier$^1$}
\address{$^1$ Nokia Bell Labs, Route de Villejust, 91620 Nozay, France\\ $^2$ Télécom Paris, 19 place Marguerite Perey, 91120 Palaiseau, France} 
\email{sterenn.guerrier@nokia-bell-labs.com}

%% Uncomment the following line to override copyright year from the default current year.
\copyrightyear{2020}

\begin{abstract}
We explore the alternatives for interrogating a fiber sensor from the polarization point of view, and demonstrate a better accuracy with dual polarization probing for coherent phi-OTDR compared with single polarization probing. %(33w here)
\end{abstract}

\section{Introduction}
Optical fiber sensors are becoming a hot topic as they %can serve in many fields, from marine environment monitoring \cite{Hartog2018_advances} to medical applications \cite{beisenova_fiber-optic_2018}. One category is the $\phi$-OTDR, capable of 
enable distributed sensing over long distances 
%and with tunable, arbitrarly small spatial resolution \textbf{[refB]}. %TODO
and thus provide low cost, lightweight sensors. 
$\phi$-OTDR exploits the retro-propagated optical field induced by the Rayleigh backscattering effect in optical fibers,%. \\ Rayleigh backscattered light effect
which is known to be polarization-dependent \cite{Deventer1993}. To overcome polarization fading effect, recent work focuses on the receiver design by using a dual-polarization coherent receiver \cite{yan_coherent_2017,martins_real_2016} so that the overall received optical power is kept constant. 
Relying on the same receiver configuration, \cite{dorize_enhancing_2018} introduces optimized probing sequences of finite length that allow a mathematically perfect channel response estimation of the sensor array. Highly sensitive measurements using this probing technique are reported in \cite{dorize_high_2019}.  

%We believe that 
As a matter of fact, the dual-polarization receiver setup is not sufficient to demonstrate a perfect channel estimation. The transmitter setup must as well be taken into account for a full understanding of a polarization insensitive sensor.
Polarization-induced phase noise was observed %in \cite{Kersey_1988_Observation} 
and further analysed in \cite{Kersey_1990_Analysis}. This effect depends on the input state of polarization (SOP), unlike polarization fading that occurs when the receiver does not detect the full state of polarization information of the backscattered signal. %at the output of the sensor depending on the receiver. 
In the following, we compare single and dual polarization interrogation schemes, considering Single (polarization) Input - Single (polarization) Output (SISO), Single Input - Multiple Output ($1\times 2$ SIMO) and Multiple Input - Multiple Output ($2\times 2$ MIMO) probing techniques, and investigate their respective issues. The study is performed using a dual-polarization backscattering model \cite{Guerrier2019_Model} and then validated by experimental results. 

\section{Polarization-sensitive sensing}
The purpose of $\phi$-OTDR is to retrieve the backscattered phase evolution from the desired fiber segments, defined by a given spatial resolution (SR), itself defined by the interrogation rate used to probe the sensor. We use the probing technique %described in 
\cite{dorize_enhancing_2018} which consists in sending coded sequences either on one or two orthogonal polarization state(s). %These coded sequences can start from single light pulses. 
At the receiver side, we consider two schemes: coherent mixers with or without polarization diversity.%The receiver can receive with or without polarization diversity. 
\subsection{Phase estimation}
\begin{itemize}
\item[--] $2\times 2$ MIMO consists in sending and receiving on two orthogonal polarizations. It allows to recover the full Jones matrix of the channel $\textbf{H} = \begin{bmatrix}
h_{xx} && h_{xy} \\ h_{yx} && h_{yy}
\end{bmatrix}$ \cite{dorize_enhancing_2018}, where the matrix coefficients are complex numbers and describe the experienced polarization crosstalk in the channel. %the $h$ are complex numbers. 
The estimated common phase is given by: %Estimated phase is: 
\begin{equation} \phi_{MIMO} = 0.5 \angle(\det{\textbf{H}}) \label{eq:mimophase} \end{equation} 
\item[--] $1\times 2$ SIMO probes a single polarization, and uses a polarization diversity receiver. Signal is recovered on both polarizations, say, if the first column of \textbf{H} is sent: %$\textbf{H'} = \begin{bmatrix} h_{xx} \\ h_{yx} \end{bmatrix}$,
 $\textbf{H'} = \begin{bmatrix}
  h_{xx} && h_{yx} \end{bmatrix}^T$,
\begin{equation}
\phi_{SIMO} = \angle(h_{xx} + h_{yx}) \label{eq:somme}
\end{equation}
Note that in \eqref{eq:somme} we also have $\phi_{SIMO}= \arctan(\dfrac{\Im_{xx} + \Im_{yx}}{\Re_{xx}+\Re_{yx}})$ with $\Re$, $\Im$ the real and imaginary parts of \textit{h} data \cite{yan_coherent_2017}. \\
\item[--] SISO probes and recovers information on one polarization (say $X$). Estimated phase is $\phi_{SISO} = \angle(h_{xx})$. 
\end{itemize}
As an event occurs along the sensor, phase variations appear in time, thus increasing the phase standard deviation locally. The main requirement for phase sensors is that no variation or ``false alarm" occurs when the fiber is not disturbed (fiber said to be in ``static mode"). The chosen metric for detection of events is the magnitude of phase standard deviation measured over the time dimension. 

\subsection{Fading and phase noise}
In this section, a backscattering model from a single mode fiber (SMF) is defined for numerical simulations. %Backscattering from a single mode fiber (SMF) is simulated. 
Rayleigh backscatterers are randomly distributed along the fiber, and so is their reflectivity (backscattered intensity). The fiber is modeled as a succession of $N$ spatial segments $i \in [1,\texttt{N}]$, each of them defined by their round-trip Jones matrix $\textbf{H}_i$. 
%The phase and intensity of the reflected optical field are determined from the distribution and characteristics of the backscatterers per segment, polarizations randomly rotate along the line and may be delayed differently \cite{Guerrier2019_Model}. 
The common phase factor and the total intensity of the reflected optical field are determined from the distribution and characteristics of the backscatterers per segment. For polarization evolution, we consider a random rotation along the segment \cite{Guerrier2019_Model}.
%As the light signal doesn't strictly go back and forth in one single fiber, but also travels a fiber between the laser and the entrance of the fiber, and then propagates from the circulator output to the coherent receiver, 
For all segments $i$, $\textbf{H}_i$ is defined as the following: 
\begin{equation}
\textbf{H}_i = A_i p_i \textbf{U}_i^T \textbf{M} \textbf{U}_i 
\end{equation}\label{eq:Hisimu} %\textbf{R}_\theta_i
where $\textbf{U}_i = \textbf{D}_ {\beta_i} \textbf{R}_{\Theta_i} \textbf{D}_{\gamma_i}$ describes a general behaviour of forward transmission in the fiber segment: $\textbf{D}$ are diagonal phase retarders and $\textbf{R}$ is a polarization rotation real matrix, all of them unitary (of determinant 1  \cite{Damask2004}). The parameters $\beta, \gamma$ are uniformly distributed over $[-\pi, \pi]$, and $\Theta_i = \arcsin{\sqrt{\xi_i }}$ with $\xi_i$ uniformly drawn in $[0,1]$. % \cite{BOYA_2003_volumes}.  %TODO keep reference only if enough space
%Then $\textbf{U}_i$ describes a general behaviour of forward transmission in the fiber segment. 
$\textbf{M}$ is a reflection matrix, modeled here as a perfect reflection (no losses, no polarization transfers) $\textbf{M}= [\begin{smallmatrix} 1 && 0 \\ 0 && -1 \end{smallmatrix}]$, $A_i$ is the attenuation along the fiber, and $p_i$ is the common phasor of the matrix derived from the simulated positions of backscatterers in the fiber \cite{Guerrier2019_Model}. 
The general shape of the Jones backscatter matrix, %where $\beta$ parameters are cancelled due to the round-trip, 
is computed as follows for all spatial segments $i$: 
\begin{equation*}\textbf{H}_i = A_i p_{i} \begin{bmatrix}
e^{j2\gamma_i}( \cos2\beta_i + j\sin 2\beta_i \cos 2\Theta_i) &&
-j\sin 2\beta_i \sin 2\Theta_i \\
-j\sin 2\beta_i \sin 2\Theta_i &&
e^{-j2\gamma_i}( \cos2\beta_i - j\sin 2\beta_i \cos 2\Theta_i)
\end{bmatrix}
\end{equation*}%\label{eq:Halphanull}

Polarization fading occurs in interferometers when two interfering light waves get orthogonal SOP so no interference occurs and no sensor phase is measured. In SISO mode, if we detect $h_{xx}$, an intensity fading will occur for 
periodic $(\Theta,\beta)$ pairs %corresponding to orthogonality between the emitted $x$-polarization and the received $x$-polarization
%$\Theta \equiv \pi/4 [\pi/2]$ (here $[.]$ stands for modulo operator) 
where no phase estimation is available. Else, no fading is brought with either SIMO or MIMO methods, there will always be a phase estimation that is stable in time: 
$\varphi_{SIMO} = \angle(p_i \times (e^{j2\gamma_i} \cos2\beta_i -j\sin 2\beta_i (e^{j2\gamma_i}\cos 2\Theta_i+\sin 2\Theta_i)))$ 
%$\varphi_{SIMO} = \angle(p_i\times(cos(2\Theta)+sin(2\Theta)e^{j2\gamma}))$ 
and $\varphi_{MIMO} = \angle(p_i)$. 
Of course, SIMO-estimated absolute phase differs from the real common phase due to rotations but, when sensing dynamic events, we are only intersted in phase variations and stability. Polarization diversity receivers have thus overcome this fading issue \cite{martins_real_2016}. 

The dependence of $\varphi_{SIMO}$ on $\Theta, \beta, \gamma$ gives rise to so-called \textit{polarization induced phase noise}. The measured sensor output phase depends on the input SOP \cite{Kersey_1990_Analysis}, and the output SOP may vary due to the birefringence in the sensor. To capture the phenomenon in a more realistic way, we model a ``TX/RX misalignment" acknowledging that a configuration where the input H and V polarizations at the transmitter are perfectly aligned with those of the receiver is very unlikely. We model the misalignment as a polarization rotation  $\textbf{R}_\theta = \begin{bmatrix}
cos{\theta} && -sin{\theta} \\ sin{\theta} && cos{\theta} \end{bmatrix} $.
 Then $\textbf{H}_i$ becomes, for any spatial segment $i$, $\textbf{H}_i = A_i p_i \textbf{U}_i^T \textbf{M} \textbf{U}_i \textbf{R}_\theta$.
Note that the input SOP includes %laser fluctuations due to phase noise and 
possible mechanical perturbations around the setup, so $\theta$ (then $\textbf{R}_\theta$) can be time-dependent. 

\section{Polarization noise simulations}
\begin{figure}[h]
\centering
\subfloat[StDv in time as a function of $\theta$, estimation on one segment, fixed $\Theta, \beta, \gamma$ \label{subfig:singleseg}]{\includegraphics[width=0.32\textwidth]{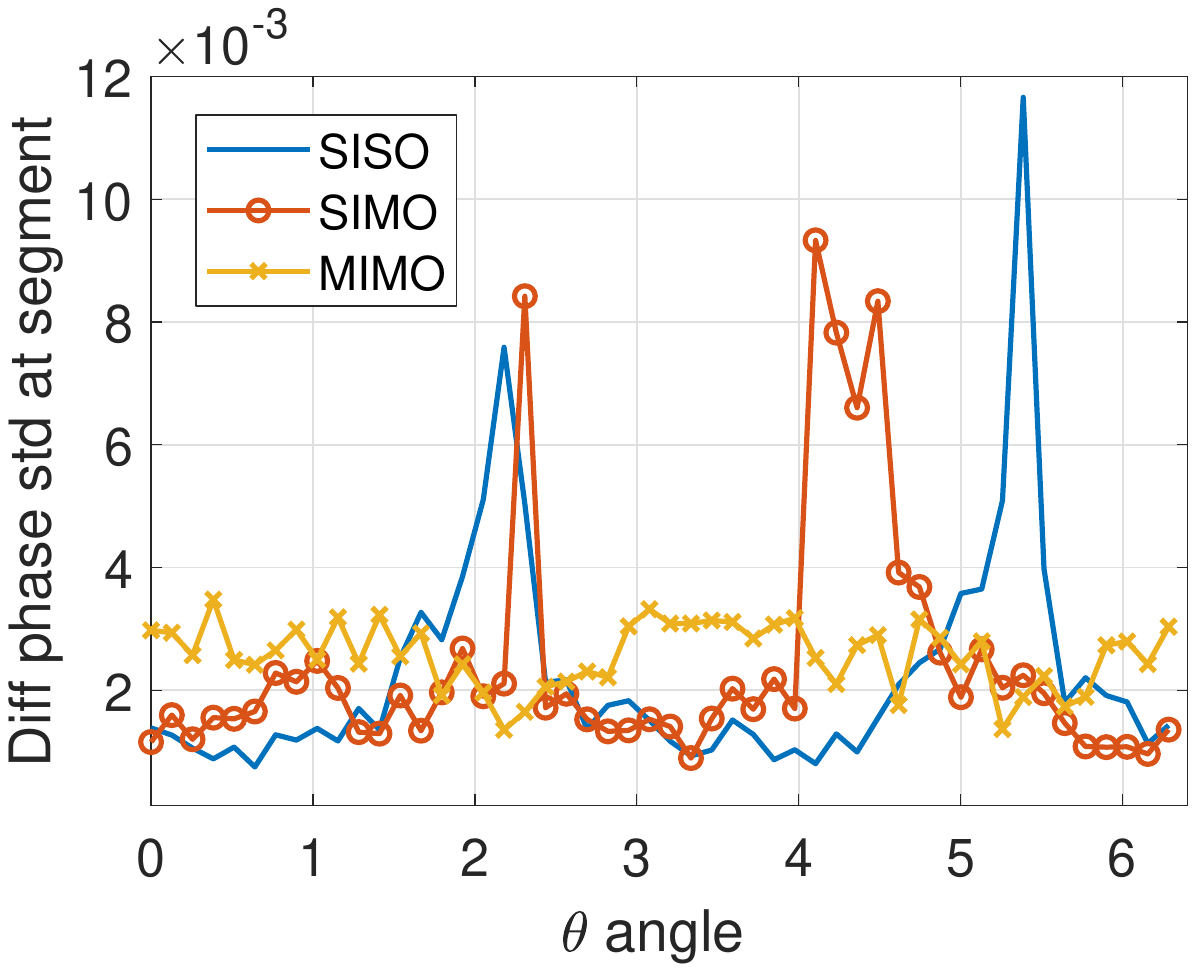}}\hspace*{0.5cm}
\subfloat[StDv in time as a function of fiber length, randomly drawn $\Theta, \beta, \theta, \gamma$ for each segment \label{subfig:1b}]{\includegraphics[width=0.32\textwidth]{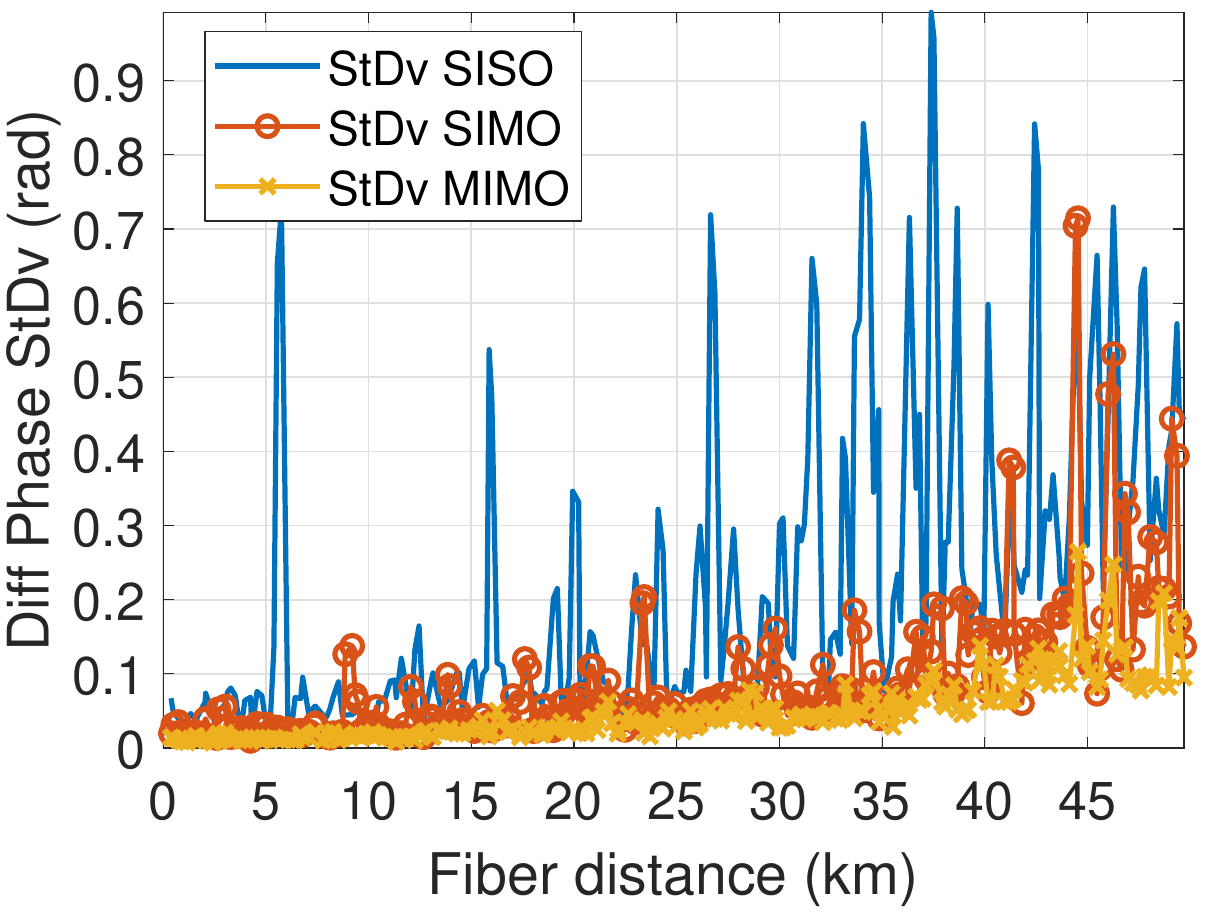}}\hspace*{0.5cm}
\subfloat[Mean StDv over 50 fibre simulations \label{subfig:meanstd}]{\includegraphics[width=0.32\textwidth]{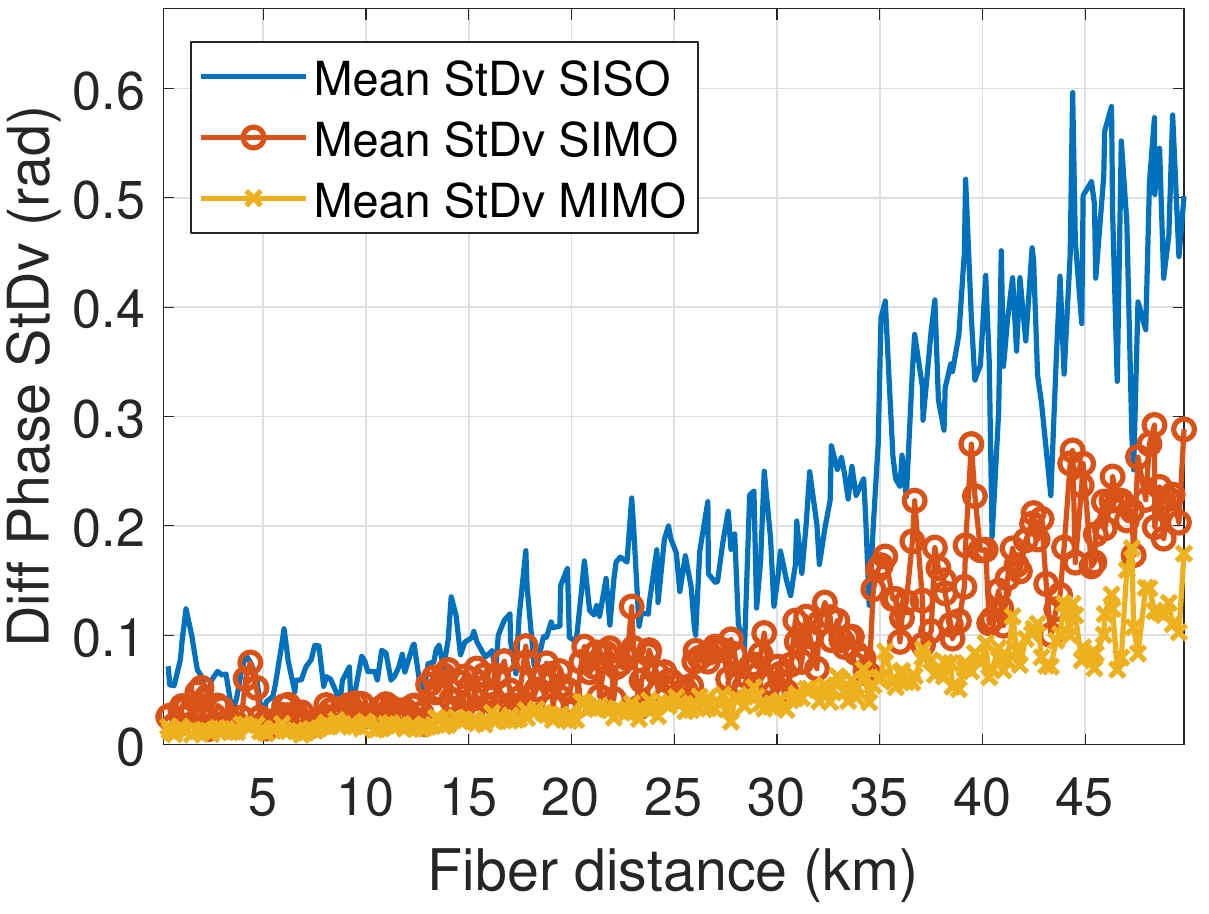}}
\caption{Standard deviation of differential phase in $\theta$ and distance, $\Delta\nu$=75Hz} 
\label{fig:stdthetadist}
\end{figure}
\noindent
We simulate Rayleigh backscattering through a fiber with a given laser phase noise due to a laser linewidth $\Delta\nu$ \cite{Guerrier2019_Model}, and observe the polarization effects on phase standard deviation in time (StDv) per fiber segment, within a time window of a few seconds, with a new phase estimation derived each 160 $\mu $s. %sb10,f=50M
For a given random  $\Theta$, some input $\theta$ may lead to higher phase variations when single-input probing techniques are used, as shown in Fig.\autoref{subfig:singleseg}. As a result, this affects the phase StDv as a function of distance (Fig.\autoref{subfig:1b}), 
where SIMO method experiences sudden variations at random fiber distances, ie. at random segment indices. 
These variations correspond to occurrences of ($\Theta$, $\theta$) pairs where the impact of input polarization induced phase noise is visible. SISO-estimated phase cumulates this effect with polarization fading due to the varying birefringence in the fiber, leading to higher values of standard deviation spread along the simulated fiber. 
MIMO probing is independent of $\theta$ and thus yields no StDv peaks along the fiber. % \\
Fig.\autoref{subfig:meanstd} highlights that on average, there is a clear hierarchy between the three studied probing methods. Particularly on long distances: after 50km, the phase noise StDv given by MIMO probing is twice smaller than the one with SIMO probing, and five times smaller than SISO. 

\begin{figure}[h]
\centering
%\subfloat[SIMO, one simulation]{\includegraphics[width=0.333\textwidth]{S48_StDImRe_Theta_SIMO}} \hspace*{1cm}
\includegraphics[width=0.42\textwidth]{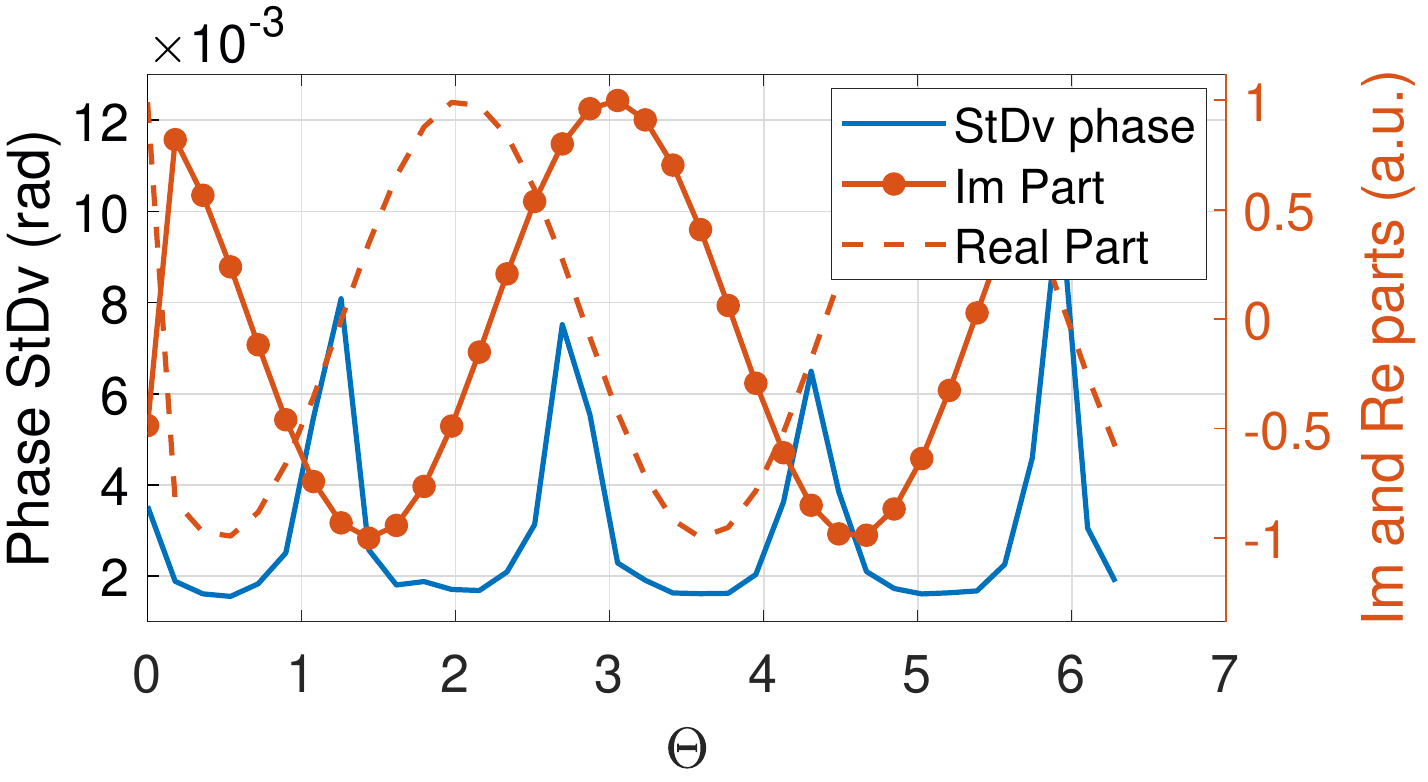}
\caption{Periodicity of $\Im(h_{xx} + h_{yx})$, $\Re(h_{xx} + h_{yx})$ and StDv over $\Theta$ (SIMO interrogation, for given $\gamma$, $\beta$, $\theta$)}\label{fig:SIMOphaseReIm}
\end{figure}

%%%%%%%%% %TODO %%%%%% Explication de l'influence du bruit de phase induit par la polar %%%%%%%%%%%%%%%%
Back to a SIMO configuration, we choose fixed $\theta$ (polarization misalignment on the transmitter side) and $(\beta,\gamma)$ (phase retardances) to follow real (denoted $\Re$) and imaginary (denoted $\Im$) parts of $h_{xx} + h_{yx}$ for different $\Theta$ (polarization rotation in a fiber segment) in \autoref{fig:SIMOphaseReIm}. The standard deviation of the phase $\angle(h_{xx} + h_{yx})$ for that segment is superimposed. 
As SIMO-estimated phase is a modulation of the common phasor $p_i$ by the polarization parameters $\Theta, \beta$ and $\gamma$ (polarization phase noise) of the fiber segment as follows :
$\varphi_{SIMO} = \angle(p_i \times (e^{j2\gamma_i} \cos2\beta_i -j\sin 2\beta_i (e^{j2\gamma_i}\cos 2\Theta_i+\sin 2\Theta_i)))$, 
% $\varphi_{SIMO} = \angle(p_i\times(cos(2\Theta)+sin(2\Theta)e^{j2\gamma}))$, 
we expect the $\Re$ and $\Im$ parts to be modulated as a function of $\Theta$. 
%%%%%%%%%%%%%%%%%%%%%%%%%%%
 \\
%The reason for StDv peaks in presence of input polarization phase noise is shown to have numerical grounds: a peak occurs each time $\Re(h_{xx} + h_{yx})$ is null as the phase is calculated via $\arctan\dfrac{\Im(h_{xx} + h_{yx})}{\Re(h_{xx} + h_{yx})} $, this appears as a mere division-by-zero issue: increased instability around undefined points.  
That modulation as a function of $\Theta$ not only changes the value of the phase estimation ($\Im$ part), but also yields phase StDv peaks when $\Re(h_{xx} + h_{yx})$ is close to zero, inducing a fading effect. %Indeed, it is known that if no or few power reaches the receiver, then the phase estimation will be uncertain: fading occurs. 
As this ``input polarization fading" occurs periodically as a function of $\Theta$ (as well as $\gamma$ and $ \beta$, not displayed for simplicity) in the fibre and since there are no means for controlling such an intrinsic parameter of the fiber sensor, % the SIMO interrogation method is definitely sensitive to polarization-induced phase noise
SIMO interrogation method is thus limited by polarization-induced phase noise, even though we should be capable of retrieving a phase at all times from a polarization insensitive receiver. 
%TODO point sur les solutions proposées par Kersey et al : depolarized ligth 
In non-coherent interferometric setups, polarization induced phase noise appears as phase fluctuations and thus visibility fluctuations at the receiver side \cite{Kersey_1990_Analysis}. In that case, the standard solution to overcome this issue is to use depolarized light at the transmitter. We investigate here 
an alternate solution, based on joint probing of two polarizations: MIMO probing method is immune to polarization phase noise, thus experiences potential phase StDv peaks only when the backscattered intensity of a segment is low. 
\section{Experimental validation}
\begin{figure}[bh]
\centering
\includegraphics[width=0.9\textwidth]{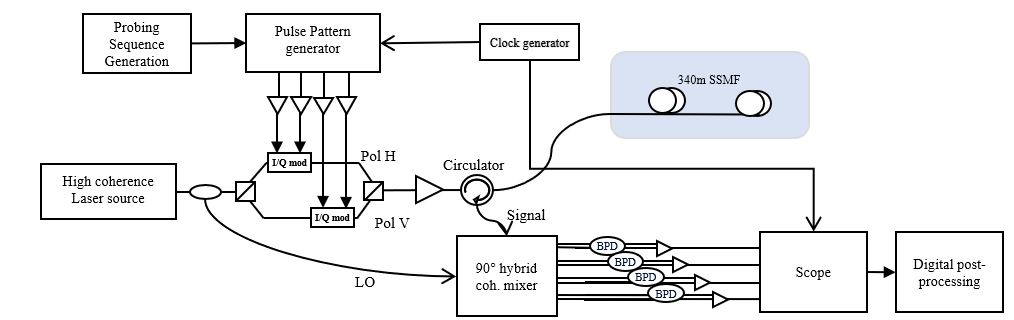}
\caption{Experimental setup}\label{fig:expsetup}
\end{figure}
\begin{figure}[h]
\centering
\includegraphics[width=0.38\textwidth]{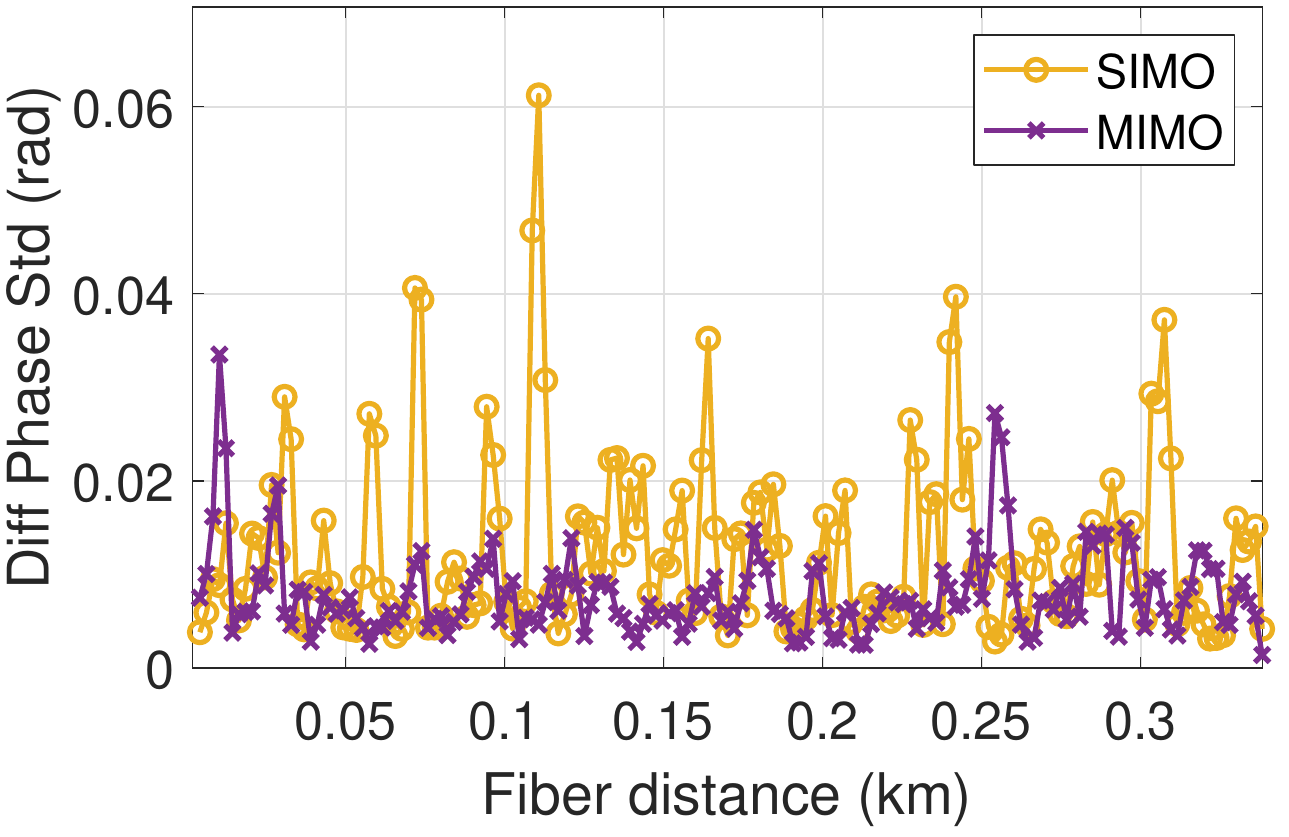}
\caption{SIMO and MIMO optical phase standard deviations as a function of fiber length, 340m SSMF} \label{fig:expresult}
\end{figure}

 \noindent
We perform measurements on a 340m-long standard single mode fiber (SSMF). The fiber experiences no strain so to stay in ``static" conditions, as it was for the simulations. The experimental setup is shown in \autoref{fig:expsetup}. 
Both measurements couldn't be performed simultaneously but were made successively so to keep the same/closest as possible conditions for the setup. For the SIMO measurement, the modulator modulates complementary binary sequences on one polarization only. Then, the second polarization is modulated too, with orthogonal sequences, and the MIMO measurement is performed. The measured standard deviation (StDv) of the phase is plotted \autoref{fig:expresult}. 
We observe StDv peaks for both SIMO and MIMO probing due to low backscattered power from the fiber segments. The peaks are not perfectly aligned due to mere synchronization issues. Both StDv are low since no perturbation is applied to the sensing fiber, however we observe a high number of StDv peaks for SIMO whereas MIMO has only two (attributed to low intensity backscatter). 
The experimental result shows a better phase stability using MIMO probing, thus confirming our modeling results.
%One extra line for other exp comments
\section{Conclusions}
We pointed out the existence of two different polarization-related impairments in interferometric sensors and observed their influence through simulations. %simulated their effects. 
Simulation allowed to isolate some effects so to separate the contribution of independent disturbances which usually occur together in experimental setups. This study allows to claim that the presence of a polarization diversity receiver or any other technique to mitigate polarization fading is a necessary condition but not sufficient to overcome all polarization effects occurring along the sensor. Dual-polarization probing of the fiber sensor makes it insensitive to the input polarization fluctuation effects, thus dividing the phase noise by two on long distances. It appears as a promising complement to the polarization diversity receiver in a coherent-$\phi$-OTDR setup.   

%This study also brings in some leads towards an enhancement of our MIMO interrogator, on which the criteria of $\Re$ part can have residual influence. 

%\section{Main Text}
%
%\subsection{Required Elements}
%All PDF submissions must contain the following items in order to be published:
%
%\begin{enumerate}
%\item Complete title
%\item Complete listing of all authors and their affiliations
%\item Self-contained abstract (indexers such as Google Scholar will not index papers that do not contain abstracts)
%\item Appropriate copyright statement following the abstract. By default, the copyright statement will appear as \number\year \hskip.05in The Author(s). If needed, the default statement can be suppressed by use of the \verb+{abstract*}+ environment.
%\item Permission and attribution for any trademarked or copyright images. Note that images of people or images owned or trademarked by other entities (including well-known logo's or cartoon characters for example) will also require official written permission.
%\item 4-page Summary: the author must include all text, including the 35-word abstract, title, authors, equations, tables, photographs, drawings, figures, and references. The text may be typed either single-spaced or double-spaced. Refrain from use of asterisk, job descriptions, or footnotes.
%\end{enumerate}

\vfill
\bibliographystyle{osajnl}
\bibliography{bibliographie}\label{sec:bib}
\end{document}